\documentclass[fleqn,usenatbib]{mnras}

\usepackage{newtxtext,newtxmath}

\usepackage[T1]{fontenc}
\usepackage{ae,aecompl}

\usepackage{graphicx}	
\usepackage{amsmath}	
\usepackage{amssymb}	

\newcommand{\msun}{\mathrm{M}_\odot}

\title[White-Dwarf Convection]{The Onset of Convective Coupling and Freezing in the White Dwarfs of 47 Tucanae}

\author[A. Obertas et al.]{Alysa Obertas,$^{1}$\thanks{Contact e-mail:\href{mailto:obertas@astro.utoronto.ca}{obertas@astro.utoronto.ca}}\thanks{
Current addresses:
Department of Astronomy and Astrophysics, University of Toronto, 50 St. George Street, Toronto, ON M5S 3H4, Canada; and
Canadian Institute for Theoretical Astrophysics, University of Toronto, 60 St. George Street, Toronto, ON M5S 3H8, Canada}
Ilaria Caiazzo,$^{1}$
Jeremy Heyl,$^{1}$
Harvey Richer$^{1}$,
Jason Kalirai$^{2,3}$,
\newauthor
and Pier-Emmanuel Tremblay$^{4}$
\\
$^{1}$Department of Physics and Astronomy, University of British Columbia, 6224 Agricultural Road, Vancouver, BC V6T 1Z1, Canada\\
$^{2}$Space Telescope Science Institute, 3700 San Martin Drive, Baltimore MD 21218 \\
$^{3}$Center for Astrophysical Sciences, Johns Hopkins University, Baltimore MD, 21218\\
$^{4}$Department of Physics, University of Warwick, Coventry CV4 7AL, UK
}

\date{Accepted XXX. Received YYY; in original form ZZZ}

\pubyear{2017}

\begin{document}
\label{firstpage}
\pagerange{\pageref{firstpage}--\pageref{lastpage}}
\maketitle

\begin{abstract}
Using images from the Hubble Space Telescope Advanced Camera for Surveys, we measure the rate of cooling of white dwarfs in the globular cluster 47~Tucanae and compare it to modelled cooling curves. We examine the effects of the outer convective envelope reaching the nearly isothermal degenerate core and the release of latent heat during core crystallisation on the white dwarf cooling rates. For white dwarfs typical of 47~Tuc, the onset of these effects occur at similar times. The latent heat released during crystallisation is a small heat source. In contrast, the heat reservoir of the degenerate core is substantially larger. When the convective envelope reaches the nearly isothermal interior of the white dwarf, the star becomes brighter than it would be in the absence of this effect. Our modelled cooling curves that include this convective coupling closely match the observed luminosity function of the white dwarfs in 47 Tuc. 
\end{abstract}




\section{Introduction}
\label{sec:intro}

White dwarfs (WDs) are the remnants of stars with initial masses up to $8~\msun$ and are the fate of at least 95\% of stars. Unlike their progenitors, nuclear fusion reactions no longer occur in their cores. Young WDs, initially very hot and bright, consequently cool down and become fainter as they age. After cooling down for some time, WD cores begin to crystallise. Latent heat is released during the crystallisation phase transition \citep{1967AJ.....72R.834V,1968ApJ...151..227V}, slowing down cooling in the atmosphere and the change in observed effective temperature.  \citet{1968ApJ...151..227V} argued that the crystallisation would occur when the core temperature was about $6\times 10^6$~K and the luminosity about $10^{-3.3}~\mathrm{L}_\odot$.

Crystallisation plays a critical role in the WD cooling process and is therefore a necessary inclusion in WD cooling models.  This is important when considering applications of WD cooling for old stellar populations such as those in the Milky Way bulge, halo, and globular clusters.  Globular clusters are of particular interest as they are some of the oldest structures in the Milky Way. Understanding their formation time and subsequent evolution can be directly tied to the accretion history of the Galactic halo.

The study of WD crystallisation began in the 1960s, although it was not until the 1990s that it was realised that observations of pulsating WDs could be used to find evidence of crystallisation \citep{1995BaltA...4..129W}. Techniques in asteroseismology allowed for this breakthrough. Gravito-acoustic oscillations in WDs enable internal structure to be probed through observations \citep{2004ApJ...605L.133M}. There has been limited research conducted to find evidence of WD crystallisation in stellar populations. \citet{2009ApJ...693L...6W} argued that they found evidence on crystallisation in the luminosity function of the white dwarfs in NGC~6397. The current WD cooling models, which are being applied to determine ages of stellar populations, include the effects of core crystallisation. For this reason, it is important to search for evidence of core crystallisation in these populations to verify the cooling models or adjust them accordingly.

At approximately the same time that \citet{1967AJ.....72R.834V} proposed that white dwarfs would freeze, \citet{1968Ap&SS...2..375B} argued that their outer layers, the envelopes, would be convective. \citet{1990ApJS...72..335T} argued that for a 0.6~solar mass white dwarf, the convective layer would break through to the degenerate core of the white dwarf when the luminosity was about $10^{-3.7} \mathrm{L}_\odot$, about the same luminosity as found by \citet{1968ApJ...151..227V} for the onset of crystallisation.  \citet{2001PASP..113..409F} in a classic review paper argue that the convective coupling of the core to the surface changes the cooling evolution of the star, resulting in its appearing initially more luminous than it would have been had the energy continued to flow radiatively and that the increase in the luminosity from the convective coupling dominates over the additional latent heat \citep[see also][]{2015ApJ...799..142T}. 

47 Tucanae (47 Tuc) is a globular cluster at a distance of 4.7 kpc \citep{2012AJ....143...50W} in the constellation Tucana. 47 Tuc is a good candidate to search for WD core crystallisation and convection, because it is an old globular cluster (age $\sim 10$~Gyr, \citealt{2013Natur.500...51H}) with a large population of WDs.  The typical mass of these WDs is about $0.53~\msun$ \citep{2009ApJ...705..408K,2012Natur.488..684K}, a bit lower than that discussed by \citet{1990ApJS...72..335T}.   In this case, the critical luminosity where the centre of the star freezes is somewhat lower \citep{1968ApJ...151..227V}, and the luminosity where convection begins to be important is somewhat higher \citep{1990ApJS...72..335T}; therefore, we expect to see an interplay of freezing and convection in the white dwarfs of 47~Tucanae.

\section{Observations}
\label{sec:obs}

\begin{figure}
\includegraphics[width=\columnwidth,clip,trim=0 0.5in 0 0.5in]{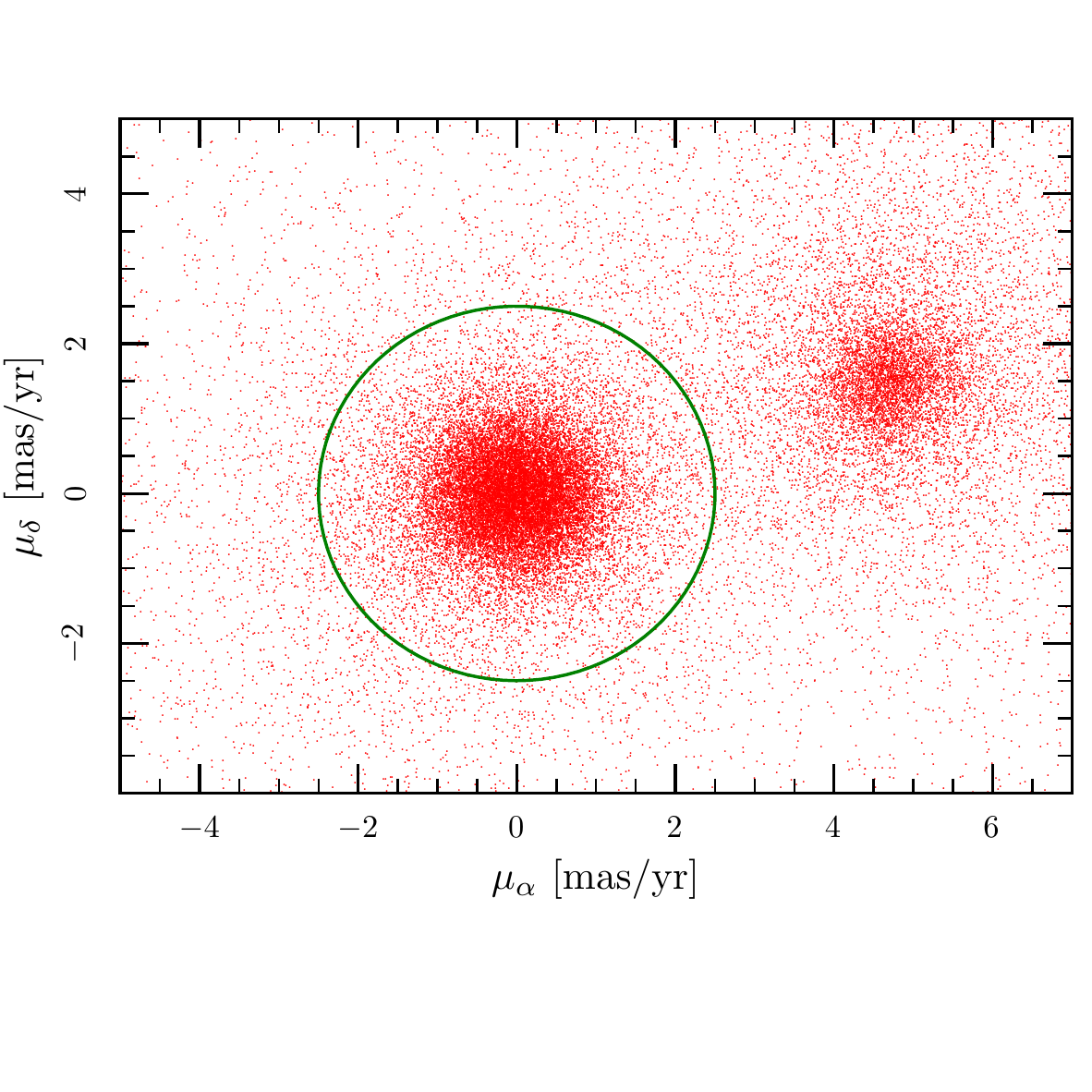}
\includegraphics[width=\columnwidth]{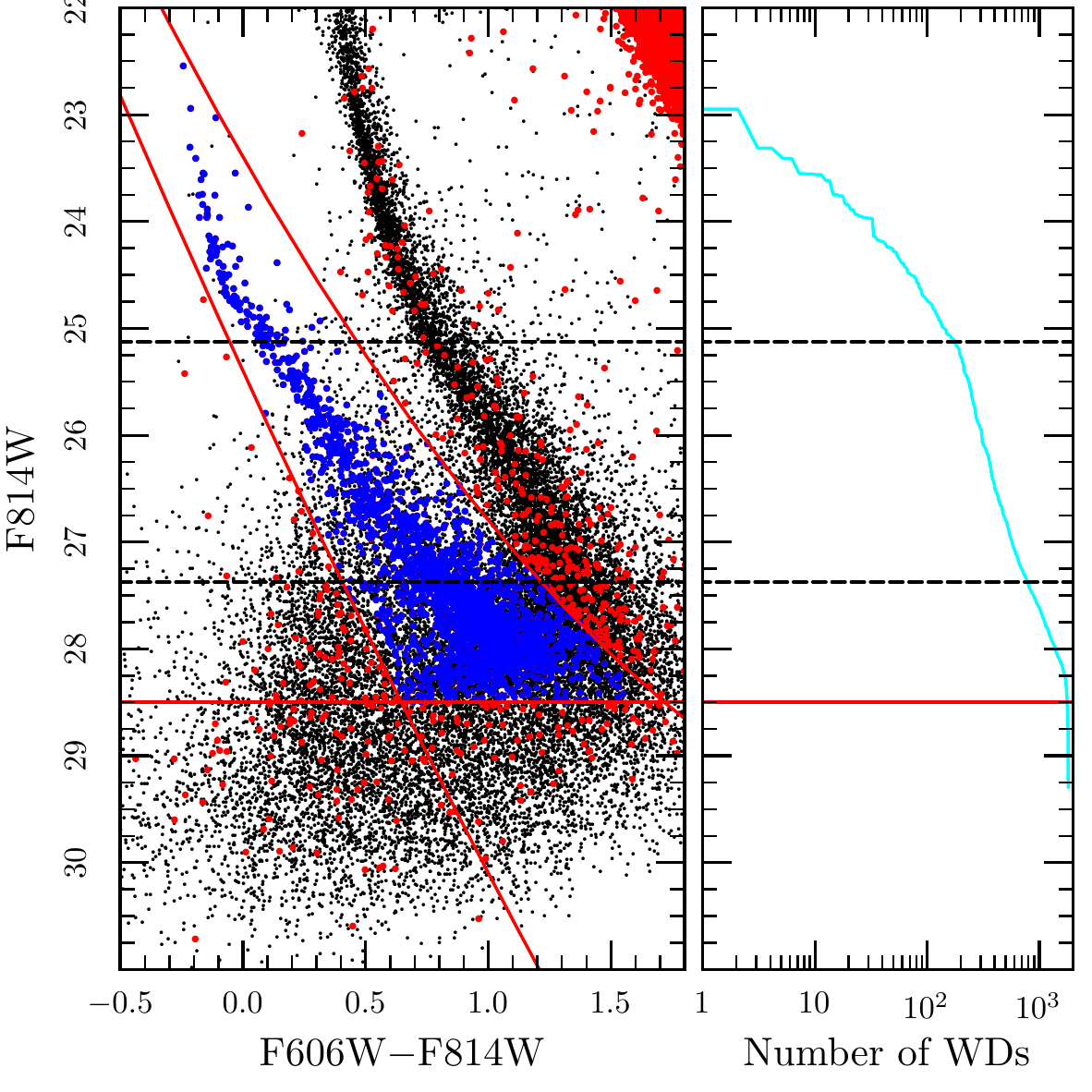}
\caption{Upper panel: the proper-motion diagram of our dataset.  Lower panel: the proper-motion-cleaned colour-magnitude diagram of our field in 47~Tucanae focusing on the white dwarf region. The black points denote stars that lie outside the green circle in the upper panel.  These are mainly stars in the SMC that lies in the background of 47 Tuc. Our sample of white dwarfs are the blue dots within the region.  Above the upper black dashed line, the cooling is dominated by neutrinos, and below the lower dashed line, there is a kink in the luminosity function of white dwarfs (see Fig.~\ref{fig:f814_fr2}). The right panel depicts the number counts.  Note the changes in slopes at the magnitudes indicated.}
\label{fig:cmd}
\end{figure}

In this study, we use the 47 Tucanae ACS field described by \citet{1538-3881-143-1-11} (see Fig.~\ref{fig:cmd}). The field is 6.7 arcminutes from the centre of the cluster. The observations of this field with Hubble's Advanced Camera for Surveys comprise over 150~ks of exposure in both F606W and F814W.  This is one of the deepest observations of a globular cluster; they probe deep enough to study the entirety of the white dwarf cooling sequence. \citet{Rich1347tuc} describe the reduction procedures in further detail.  The proper motion of each star is measured relative to the mean motion of other stars in the cluster, so these observations are not directly sensitive to the rotation of the cluster \citep{2003AJ....126..772A}.   We determine the proper motions from the combined images; this means that although the proper motions are less accurate, we can obtain proper motions for all of the detected stars and exclude stars that do not move with approximately the mean proper motion of the cluster (see the upper panel of Fig.~\ref{fig:cmd}).  We use the completeness calculations performed in \citet{1538-3881-143-1-11} (see their Fig.~9) to estimate the properties of the underlying distribution of white dwarfs.

Our sample of white dwarfs extends to nearly the fifty-percent completeness threshold of F606W=29.75 and F814W=28.75.  As most of our white dwarfs are bluer than F606W-F814W=1, we will use the completeness rate in F814W to determine the cumulative distribution of white dwarfs as a function of F814W magnitude as follows,
\begin{equation}
C(<m) = \sum_{m_i\leq m} \frac{1}{C_i}
\label{eq:0}
\end{equation}
where $m_i$ is the apparent magnitude of each star in the sample and $C_i$ is the completeness rate for a star of that magnitude. 

\section{White-Dwarf Evolution}
\label{sec:wd_evol}

We calculate the full white dwarf evolution using MESA stellar models \citep{2011ApJS..192....3P} and the DA stellar atmospheres of \citet{2006AJ....132.1221H}, \citet{2006ApJ...651L.137K} and \citet{2011ApJ...730..128T}\footnote{http://www.astro.umontreal.ca/~bergeron/CoolingModels} to determine the F814W flux of the white dwarfs as a function of age.  We used MESA revision 7624 starting with the inlist file in the test suite called ``make\_co\_wd''.  We used an initial mass of 0.974 solar masses which yields a formation time of the white dwarf of 8.6~Gyr, so the white dwarfs start to freeze about 10~Gyr after the birth of the progenitor star, approximately the age of the cluster today.  We set the metallicity to $Z=0.004$ corresponding to $[\mathrm{Fe/H}]=-0.75$.  The properties of the white dwarf are relatively insensitive to these choices.  We set the  $\eta$ parameter for the \citet{1975MSRSL...8..369R} wind on the red-giant branch to 0.1 and for the \citet{1995A&A...297..727B} wind on the asymptotic giant branch to 0.15. This results in little mass loss on the red giant branch as we found from our observations in the core of 47~Tuc \citep{2015ApJ...810..127H}.  Varying the wind parameter during the red giant phase by a large factor can change the mass of the white dwarf by a few hundredths of a solar mass. On the other hand, changes to the wind during the AGB changes the masses of the thin hydrogen and helium layers that lie above the carbon-oxygen core of the white dwarf.

\begin{figure}
\includegraphics[width=\columnwidth]{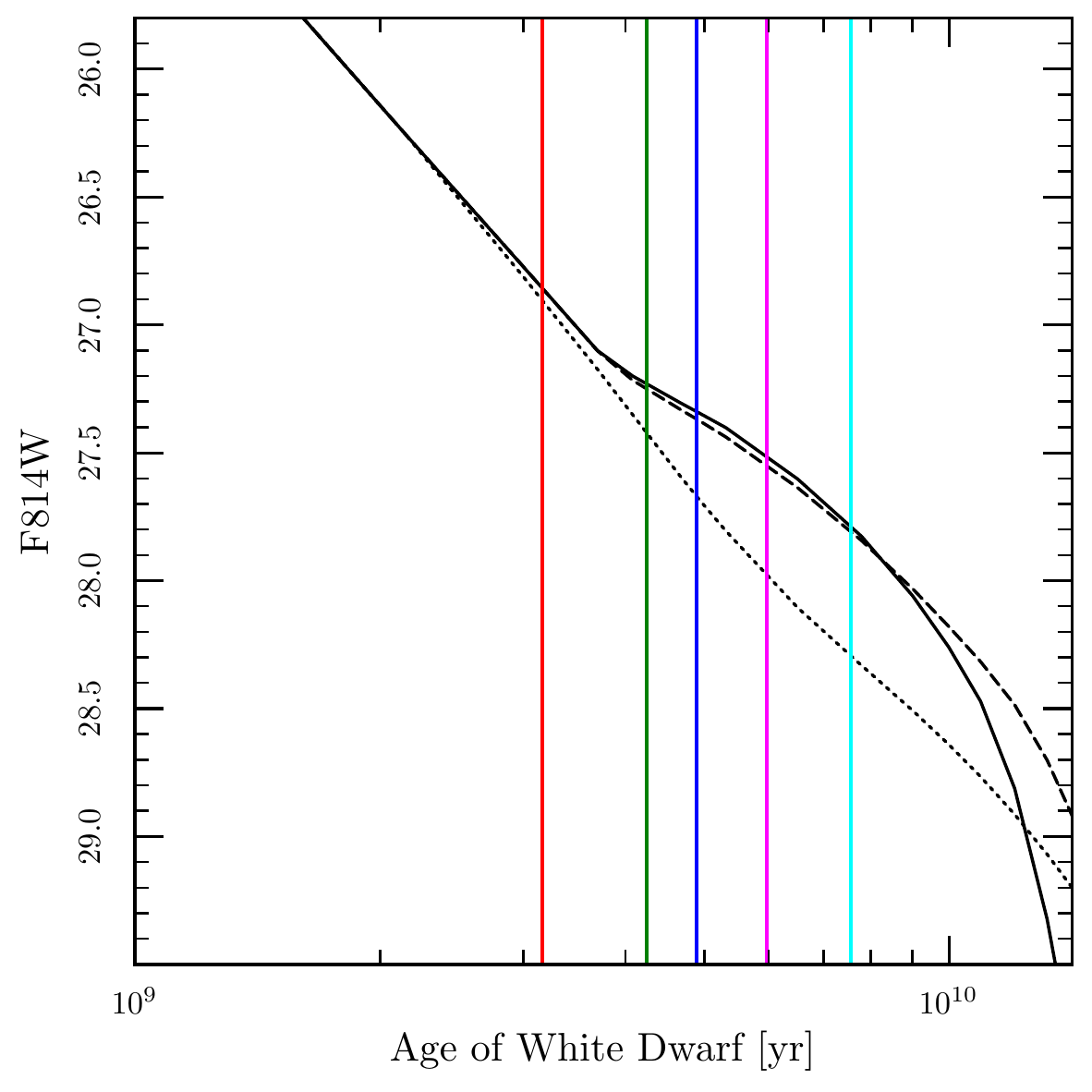}
\caption{The simulated F814W magnitude as a function of time for the white dwarfs like those in 47~Tucanae.  The solid black curve depicts the full evolutionary model.  The dashed black curve depicts a model without freezing.  The dotted black curve depicts a model without freezing or convection.  The leftmost vertical line indicates that the very centre of the star has begun to freeze, and each successive vertical line indicates that an additional tenth of a solar mass has frozen. The colours of the vertical lines corresponds to the epochs of the model interiors depicted in Figs.~\ref{fig:linear_entropy} and~\ref{fig:density_entropy2}.}
\label{fig:f814_fr}
\end{figure}

These theoretical models, of course, include both freezing and convection.   The latent heat is estimated using the equation of state of \citet{2010CoPP...50...82P}.  To determine the evolution without freezing and without convection, we divide the white dwarf into the core and envelope.  We assume that the core contains all of the heat capacity but none of the opacity and that the envelope contains all of the opacity and none of the heat capacity.  In both cases we use the output of the MESA models to establish the relationship between the central temperature and the luminosity of the star.  We evolve the star forward in both cases using the equation
\begin{equation}
c_\mathrm{WD} \frac{d T_c}{dt} = L(T_c)
\label{eq:1}
\end{equation}
where we choose the value of $c_\mathrm{WD}$ and the initial conditions to smoothly match the full calculation of the evolution before the onset of convection and before the core begins to freeze.  Because we are assuming the core does not freeze, $c_\mathrm{WD}$ is constant.  To look at the effect of convection in particular, we assume a functional form for the temperature-luminosity relation
\begin{equation}
L(T_c) \propto T_c^\alpha
\label{eq:2}
\end{equation}
where the constant of proportionality is degenerate with the value of $c_\mathrm{WD}$ and the value of $\alpha$ is chosen to match smoothly onto the earlier evolution calculated with MESA.  Fig.~\ref{fig:f814_fr} shows one particular evolution.   We start the evolution according to Eq.~\ref{eq:1} and~\ref{eq:2} at about 10~Gyr (2.4~Gyr after the birth of the white dwarf).  We find $\alpha\approx 3.08$, $L=3.5\times 10^{-3}\mathrm{L}_\odot$ and a core temperature of $T_c=5.0\times 10^6\mathrm{K}$ \citep[similar to][]{1990ApJS...72..335T}.  The fitted heat capacity is $c_\mathrm{WD} \approx 0.15 k_B \msun / m_p$ which is about $2 k_B$ per particle, appropriate for a liquid.

\begin{figure}
\includegraphics[width=\columnwidth]{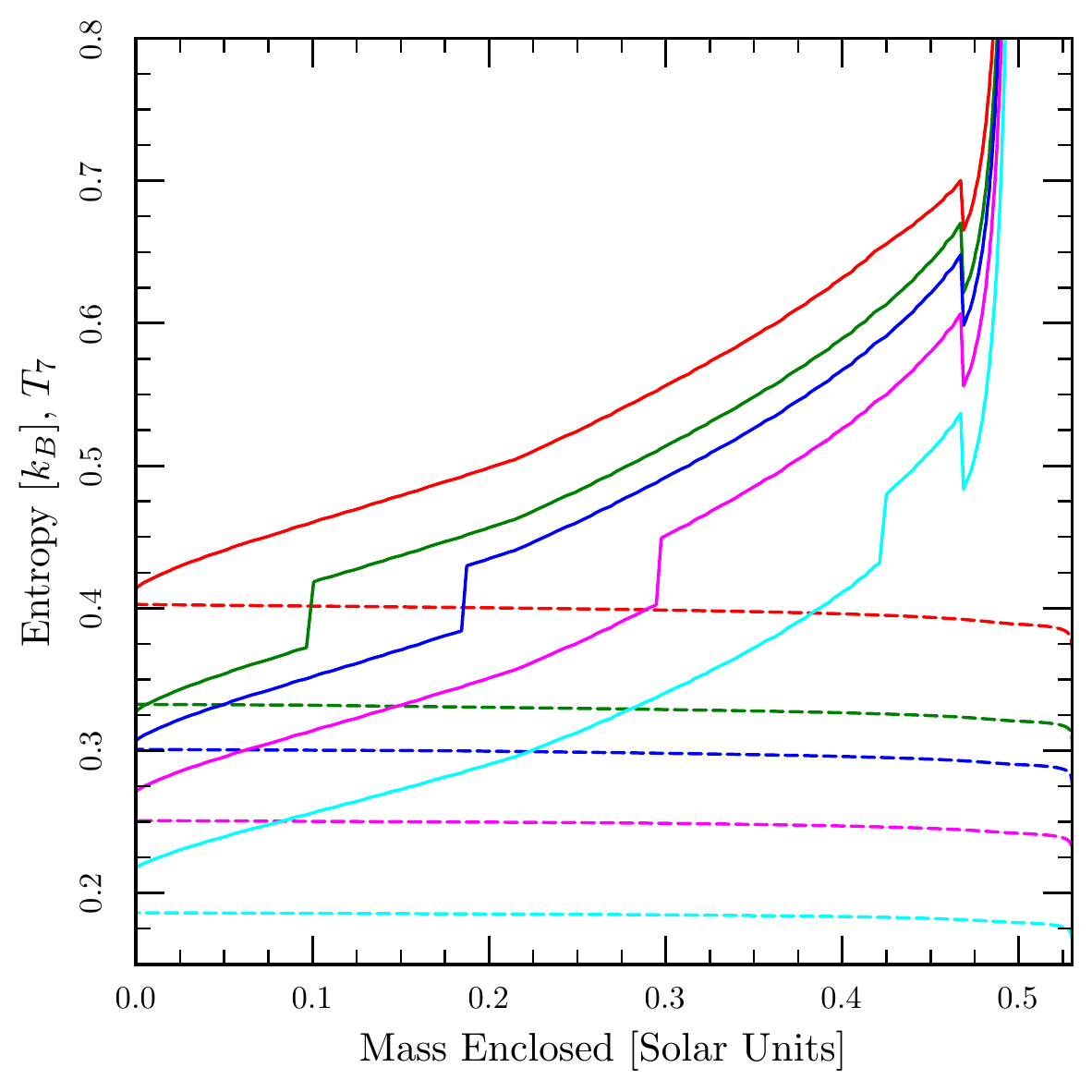}
\caption{The entropy (solid) and temperature (dashed) as a function of enclosed mass for the model interiors at the five epochs indicated by coloured vertical lines in Fig.~\ref{fig:f814_fr}. The jump in entropy near the core of the star is the release of latent heat of fusion.  The jump in entropy near the surface of the star at large enclosed mass reflects the strong composition gradient at the top of the carbon-oxygen core of the white dwarf.}
\label{fig:linear_entropy}
\end{figure}

\begin{figure}
\includegraphics[width=\columnwidth]{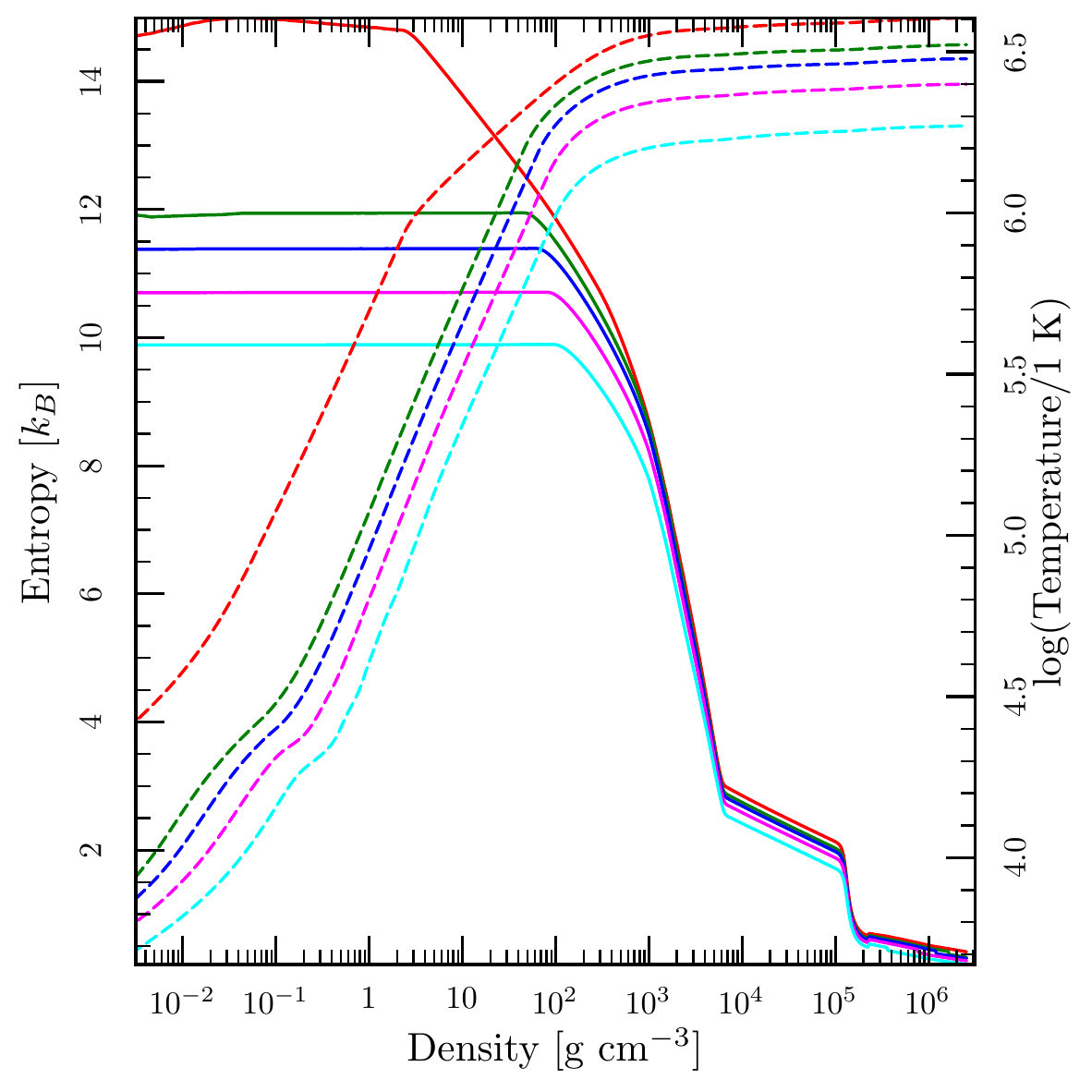}
\caption{The entropy (solid) and temperature (dashed) as a function of density for the model interiors at the five epochs indicated in Fig.~\ref{fig:f814_fr}. A hallmark of a convective region is the increase in entropy with density.  Low-entropy material resting above high-entropy material is convectively unstable.}
\label{fig:density_entropy2}
\end{figure}

The solid black curve in Fig.~\ref{fig:f814_fr} is the complete evolution of the white dwarf including both freezing and convection. We have to assume a distance to the cluster (assuming a different distance shifts the model curves vertically).  We take the distance to the cluster of 4.7~kpc \citep[][using the location of white dwarfs on the CMD]{2012AJ....143...50W} and an interstellar reddening of $E(B-V)=0.04$ \citep{2007A&A...476..243S} that translates to $A_\mathrm{F814W}=0.075$ \citep{2005PASP..117.1049S}. The leftmost vertical line indicates when the centre of the star begins to freeze and each subsequent vertical line indicates that an additional $0.1\msun$ has frozen from the centre outward.  The colours of the vertical lines correspond to the epochs of the model interiors depicted in Figs.~\ref{fig:linear_entropy} and~\ref{fig:density_entropy2}. Typically one tenth of a solar mass freezes per billion years.  Fig.~\ref{fig:linear_entropy} shows the run of entropy and temperature within the star.  The typical freezing temperature is about $3\times 10^6 \mathrm{K}$, and the freezing corresponds to a decrease in entropy of about $k_B/20$. Combining these results yields the the rate of release of latent heat which is about $1.5 \times 10^{29}~\mathrm{erg~s}^{-1}$ or $4\times 10^{-5} \mathrm{L}_\odot$.  The total luminosity from the surface at this time is about $4\times 10^{-4} \mathrm{L}_\odot$, ten times larger, so the latent heat only makes a minor contribution to the power budget.  This is reflected in the fact that the solid curve lies only slightly above the dashed curve while the freezing is happening.

The dashed black curve in Fig.~\ref{fig:f814_fr} shows the evolution where the interior of the star does not freeze, using Eq.~\ref{eq:1}. The evolution of the star with freezing is nearly identical to the one without freezing \citep[reproducing the conclusions of][]{2001PASP..113..409F}; it is initially slightly brighter due to the latent heat.   Once nearly the entire star is frozen, the flux of the frozen star begins to decrease more rapidly than for the star that does not freeze.  This effect is not due to the release of latent heat because this would be an additional energy source and would increase the flux (furthermore, all of the latent heat has already been released by this time).  Rather, this more rapid cooling results from the decrease in the specific heat of the solid phase relative to the liquid phase. For the unfrozen model assumed that the specific heat in the liquid phase was constant.  In the solid phase there are two sources of specific heat.  At high temperatures phonons dominate with $c_{WD} \propto T_c^3$, and at lower temperatures thermally excited electrons at the top of the Fermi sea dominate with $c_{WD} \propto T_c$.  In both cases the temperature dependence of the specific heat will cause the flux to decrease more rapidly.  The hallmark of freezing is not an increase in the luminosity at a given time due to the release of latent heat but rather a decrease in luminosity due to the decrease in the specific heat after the star has frozen.

The dotted black curve in Fig.~\ref{fig:f814_fr} shows the evolution without freezing and with the assumption that the radiative diffusion continues to dominate the heat transport through the envelope of the star (Eq.~\ref{eq:2}).  In this case, there is no kink in the cooling curve.  Indeed, the kink in the complete cooling curve (black solid line) happens at a time that is only coincidentally about when the star begins to freeze. The kink does not results not from freezing: it is due to the growth of convection in the surface layers of the star until the convective region reaches all of the way to the degenerate core as shown in Fig.~\ref{fig:density_entropy2}.  Although at the epochs depicted by the red and green curves there is convection very close to the surface, the convective layer is too shallow to affect the outgoing radiation dramatically.  By the epoch of the blue model, the convective layer extends all the way to the nearly isothermal core, and the heat transport is now dominated by convection.  The luminosity is much larger than it would have been otherwise, resulting in a bump in the cooling curve.

\section{Results}
\label{sec:results}

\begin{figure}
\includegraphics[width=\columnwidth]{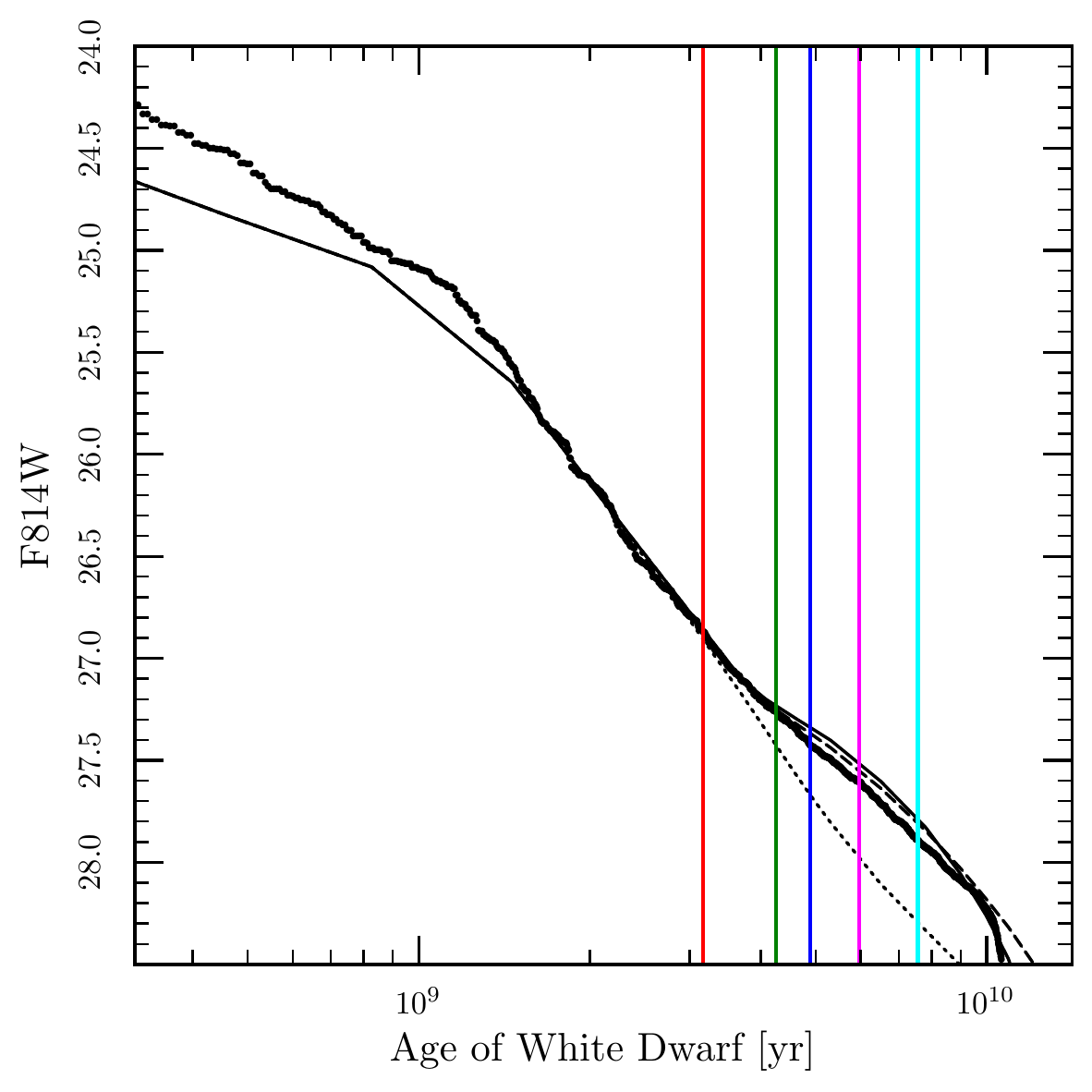}
\caption{The simulated (line styles as in Fig.~\ref{fig:f814_fr} and observed F814W magnitudes (black points) as a function of time for the white dwarfs like those in 47~Tucanae.  Relative to Fig.~\ref{fig:f814_fr} it also depicts the early neutrino-dominated evolution.}
\label{fig:f814_fr2}
\end{figure}

Fig.~\ref{fig:f814_fr2} depicts the empirical cooling curve of the white dwarfs in the field \citep{2012ApJ...760...78G} against the fiducial models discussed so far.  To convert the number of white dwarfs to an age we have to divide by the birthrate of white dwarfs in the field which we estimate to be one per 6~Myr (changing this value shifts the black points horizontally).  \citet{2012ApJ...760...78G} obtain this estimate from counting the red giant stars in the field.  Because both the observed cooling curve and the theoretical models exhibit three changes in slope, the white dwarf luminosity function itself is sufficient to determine these values.  The $\mathrm{F814W}< 25.2$ region is where cooling is dominated by neutrino emission, and the $25.2 < \mathrm{F814W} < 27.2$ region is thought to be where cooling is dominated by loss of thermal kinetic energy of ions in the core through a radiative process. The increase in slope at $\mathrm{F814W} = 27.2$ coincides approximately with the beginning of the freezing of the core and with the onset of convection at the surface.   Our total sample of white dwarfs numbers about 1,433, so the change in slope in the cumulative distribution at this magnitude is highly significant with $p<10^{-15}$ by a Kolmorgorov-Smirnov test for the null hypothesis that there is no change in slope at this magnitude.

Furthermore, from the theoretical investigation presented in \S~\ref{sec:wd_evol}, we see that the observed cooling curve is consistent with or without the freezing of the core (the solid and dashed black curves in Fig.~\ref{fig:f814_fr2}), so this bump is a hallmark of the convective on the surface, not freezing per se.  As we look to even fainter magnitudes, the agreement between models and the data remains good and the data agree somewhat better with the frozen evolution (solid black curve), yielding a hint that indeed the white dwarfs have frozen. However, it is difficult to be conclusive. First, we are looking further back in time in the history of the cluster, when the birth-rate of white dwarfs might have been different than the value today. However, even 10~Gyr ago, when the cluster was just about 1~Gyr old, the turn-off mass was just twice what it is today, so it's safe to assume that the birth rate was not dramatically different.  If the birth rate was higher then, we would overestimate the age of the white dwarfs in our sample.  Second, if the completeness rate at these fainter magnitudes is higher than what we have estimated, we would also overestimate the age of these faint white dwarfs.  Both of these effects would move the black points (the empirical cooling curve) to the right.

A second more subtle discrepancy lies near the location of the kink itself.  First, the theoretical kink is much sharper than that observed.  At the flux that characterizes the kink, the errors in the flux are typically 0.3~magnitudes, so the magnitude uncertainties could easily smooth out the kink.   However, at slightly fainter magnitudes, the model is brighter than the data at a given age.  Magnitude errors will typically increase the number of brighter stars observed because there are many more fainter stars (Malmquist bias), which would result in the empirical cooling curve lying to the right of the theoretical one as it does near the kink.  The expected size of the kink depends on the thickness of the hydrogen layer on the white dwarf \citep{1990ApJS...72..335T}, so this indicates that possibly the hydrogen layers on the white dwarfs of 47~Tuc are slightly thinner than $1.35\times 10^{-4} \msun$, the value in our white-dwarf cooling model. A second possibility is that a fraction of the white dwarfs in the cluster are DB stars and therefore lack a sufficiently thick hydrogen layer to allow for the convective coupling.  A mixture of these two populations would result in a less dramatic bump from convective coupling.  However, spectroscopic studies in other globular clusters \citep{2004A&A...420..515M,2009ApJ...705..398D} have only found DA white dwarfs and no DB stars, so this possibility is unlikely.

At the brightest magnitudes there is an apparent discrepancy between the predicted neutrino cooling curve and the observed empirical luminosity function.   \citet{2012ApJ...760...78G} noted this discrepancy in a larger study of the white dwarfs of 47~Tuc that included this field. However, the follow-up study \citep{2016ApJ...821...27G} that focussed on the youngest white dwarfs in the core of 47~Tuc and studied them in the ultraviolet where the bulk of the flux emerges found that the observed distribution of white dwarfs agrees well with the expectations from the cooling models; therefore, we would argue that this difference results from uncertainties in the emission in the near-infrared from these relatively hot white dwarfs. 

\section{Conclusions}
\label{sec:conclus}

Both the observed and theoretical cooling show changes in slopes, or the dominant cooling physics, occurring at the same magnitudes. In particular, the change in slope around F814W~=~27.1, which corresponds to the onset of convective coupling \citep{2001PASP..113..409F}, is similar for the observed and simulated cooling curves and, in the observed cooling curve, is statistically significant ($p<10^{-15}$). This change in slope occurs at a magnitude slightly fainter than the magnitude where theory predicts that crystallisation will happen for a pure carbon core or a carbon-oxygen core. The signatures of crystallisation are more subtle than the onset of convection, primarily because the rate of freezing about $0.1 \msun \mathrm{Myr}^{-1}$ is too slow to contribute significantly to the energy budget of the star.  We see hints of freezing at fainter magnitudes where the frozen star cools faster than the unfrozen star because of the decrease in the heat capacity of the solid state relative to the liquid state. 

We find dramatic agreement between our white-dwarf cooling models and the observed white dwarfs of 47~Tucanae. \citet{1990ApJS...72..335T} argue that the size of the bump on the cooling curve depends sensitively on the thickness of the hydrogen layer, in particular on whether the hydrogen layer extends into the degenerate regime.  Hydrogen layers much thinner than $10^{-4} \msun$, typically do not exhibit the convective bump at these magnitudes; the hydrogen layer of the white dwarf model presented here is $1.35\times 10^{-4} \msun$ and slightly overpredicts the size of the bump. Therefore, we can use the observed cooling curve as further evidence of moderately thick hydrogen envelopes on the white dwarfs of 47 Tucanae. \citet{2016ApJ...821...27G} found that the best-fitting hydrogen layers ranged from $2.2\times 10^{-5}$ to $8.9 \times 10^{-5} \msun$ from fitting the cooling curve of young white dwarfs in 47~Tuc.

\citet{2009ApJ...693L...6W} argued that the upturn in the luminosity function of the white dwarfs of NGC~6397 between F814W of 26 and 27 is a signature of crystallisation.  Considering the difference in distance and reddening between 47 Tuc and NGC~6397 \citep{2013ApJ...778..104R}, the upturn that we observed at about F814W of 27.4, corresponds to F814W of 26.4 in NGC~6397, so from a white-dwarf-cooling point of view, the signature that they observe is likely to be the penetration of the outer convective layer into the degenerate region of the star as predicted by \citet{2001PASP..113..409F}, rather than freezing.  On the other hand,   \citet{2013ApJ...778..104R} argue that the ratio of the number white dwarfs to main sequence stars of the same mass increases much faster in NGC~6397 than can be accounted by cooling alone and faster than in 47~Tucanae.  They argue that this results from the arrival of white dwarfs into the field in NGC~6397 studied by \citet{2009ApJ...693L...6W} as they relax to their new masses. \citet{Heyl116397dyn} argued from observations of the proper motions of white dwarfs in the same field of NGC~6397 that the relaxation time for the stars in this region is greater than about 0.5~Gyr.  By measuring the cumulative luminosity function of the white dwarfs in NGC~6397 it may be possible to disentangle these two effects. Both the data and the theoretical models hint that the bump in the cooling curve caused by convection reaching the degenerate region of the star could be used as a standard candle in globular clusters, yielding distances independent of the physics and compositions of main-sequence stars.  It may also be possible to detect the hallmarks of the convective coupling in the older white dwarfs of the Galactic disk and use it to constrain the age of the components of the Galaxy more strongly \citep{2017AJ....153...10M,2017ApJ...837..162K}.

The current WD cooling models, that incorporate WD core crystallisation and surface convection, are accurate for the WDs in 47 Tuc through onset the crystallisation and convection, verifying their use for determining the ages and distances of globular clusters.  This powerful technique complements the standard method to use the main-sequence stars near the turn-off.  Although turn-off techniques use much brighter stars, the details of the turn-off are confounded by the composition of the stars, whereas the white dwarfs lack these complications. 

\section*{Acknowledgements}

This research is based on NASA/ESA Hubble Space Telescope observations obtained at the Space Telescope Science Institute, which is operated by the Association of Universities for Research in Astronomy Inc. under NASA contract NAS5-26555. These observations are associated with proposal GO-11677 (PI: Richer). This work was supported by the Natural Sciences and Engineering Research Council of Canada, the Canadian Foundation for Innovation, and the British Columbia Knowledge Development Fund.  It has made use of the NASA ADS and arXiv.org.





\bibliography{freeze}

\begin{thebibliography}{}
\makeatletter
\relax
\def\mn@urlcharsother{\let\do\@makeother \do\$\do\&\do\#\do\^\do\_\do\%\do\~}
\def\mn@doi{\begingroup\mn@urlcharsother \@ifnextchar [ {\mn@doi@}
  {\mn@doi@[]}}
\def\mn@doi@[#1]#2{\def\@tempa{#1}\ifx\@tempa\@empty \href
  {http://dx.doi.org/#2} {doi:#2}\else \href {http://dx.doi.org/#2} {#1}\fi
  \endgroup}
\def\mn@eprint#1#2{\mn@eprint@#1:#2::\@nil}
\def\mn@eprint@arXiv#1{\href {http://arxiv.org/abs/#1} {{\tt arXiv:#1}}}
\def\mn@eprint@dblp#1{\href {http://dblp.uni-trier.de/rec/bibtex/#1.xml}
  {dblp:#1}}
\def\mn@eprint@#1:#2:#3:#4\@nil{\def\@tempa {#1}\def\@tempb {#2}\def\@tempc
  {#3}\ifx \@tempc \@empty \let \@tempc \@tempb \let \@tempb \@tempa \fi \ifx
  \@tempb \@empty \def\@tempb {arXiv}\fi \@ifundefined
  {mn@eprint@\@tempb}{\@tempb:\@tempc}{\expandafter \expandafter \csname
  mn@eprint@\@tempb\endcsname \expandafter{\@tempc}}}

\bibitem[\protect\citeauthoryear{{Anderson} \& {King}}{{Anderson} \&
  {King}}{2003}]{2003AJ....126..772A}
{Anderson} J.,  {King} I.~R.,  2003, \mn@doi [\aj] {10.1086/376480}, \href
  {http://adsabs.harvard.edu/abs/2003AJ....126..772A} {126, 772}

\bibitem[\protect\citeauthoryear{{Bloecker}}{{Bloecker}}{1995}]{1995A&A...297..727B}
{Bloecker} T.,  1995, \aap, \href
  {http://adsabs.harvard.edu/abs/1995A%26A...297..727B} {297, 727}

\bibitem[\protect\citeauthoryear{{B{\"o}hm}}{{B{\"o}hm}}{1968}]{1968Ap&SS...2..375B}
{B{\"o}hm} K.-H.,  1968, \mn@doi [\apss] {10.1007/BF00650915}, \href
  {http://adsabs.harvard.edu/abs/1968Ap%26SS...2..375B} {2, 375}

\bibitem[\protect\citeauthoryear{{Davis}, {Richer}, {Rich}, {Reitzel}  \&
  {Kalirai}}{{Davis} et~al.}{2009}]{2009ApJ...705..398D}
{Davis} D.~S.,  {Richer} H.~B.,  {Rich} R.~M.,  {Reitzel} D.~R.,   {Kalirai}
  J.~S.,  2009, \mn@doi [\apj] {10.1088/0004-637X/705/1/398}, \href
  {http://adsabs.harvard.edu/abs/2009ApJ...705..398D} {705, 398}

\bibitem[\protect\citeauthoryear{{Fontaine}, {Brassard}  \&
  {Bergeron}}{{Fontaine} et~al.}{2001}]{2001PASP..113..409F}
{Fontaine} G.,  {Brassard} P.,   {Bergeron} P.,  2001, \mn@doi [\pasp]
  {10.1086/319535}, \href {http://adsabs.harvard.edu/abs/2001PASP..113..409F}
  {113, 409}

\bibitem[\protect\citeauthoryear{{Goldsbury} et~al.,}{{Goldsbury}
  et~al.}{2012}]{2012ApJ...760...78G}
{Goldsbury} R.,  et~al., 2012, \mn@doi [\apj] {10.1088/0004-637X/760/1/78},
  \href {http://adsabs.harvard.edu/abs/2012ApJ...760...78G} {760, 78}

\bibitem[\protect\citeauthoryear{{Goldsbury}, {Heyl}, {Richer}, {Kalirai}  \&
  {Tremblay}}{{Goldsbury} et~al.}{2016}]{2016ApJ...821...27G}
{Goldsbury} R.,  {Heyl} J.,  {Richer} H.~B.,  {Kalirai} J.~S.,   {Tremblay}
  P.~E.,  2016, \mn@doi [\apj] {10.3847/0004-637X/821/1/27}, \href
  {http://adsabs.harvard.edu/abs/2016ApJ...821...27G} {821, 27}

\bibitem[\protect\citeauthoryear{{Hansen} et~al.,}{{Hansen}
  et~al.}{2013}]{2013Natur.500...51H}
{Hansen} B.~M.~S.,  et~al., 2013, \mn@doi [\nat] {10.1038/nature12334}, \href
  {http://adsabs.harvard.edu/abs/2013Natur.500...51H} {500, 51}

\bibitem[\protect\citeauthoryear{Heyl et~al.,}{Heyl
  et~al.}{2012}]{Heyl116397dyn}
Heyl J.~S.,  et~al., 2012, \apj, 761, 51 (25 pages)

\bibitem[\protect\citeauthoryear{{Heyl}, {Kalirai}, {Richer}, {Marigo},
  {Antolini}, {Goldsbury}  \& {Parada}}{{Heyl}
  et~al.}{2015}]{2015ApJ...810..127H}
{Heyl} J.,  {Kalirai} J.,  {Richer} H.~B.,  {Marigo} P.,  {Antolini} E.,
  {Goldsbury} R.,   {Parada} J.,  2015, \mn@doi [\apj]
  {10.1088/0004-637X/810/2/127}, \href
  {http://adsabs.harvard.edu/abs/2015ApJ...810..127H} {810, 127}

\bibitem[\protect\citeauthoryear{{Holberg} \& {Bergeron}}{{Holberg} \&
  {Bergeron}}{2006}]{2006AJ....132.1221H}
{Holberg} J.~B.,  {Bergeron} P.,  2006, \mn@doi [\aj] {10.1086/505938}, \href
  {http://cdsads.u-strasbg.fr/abs/2006AJ....132.1221H} {132, 1221}

\bibitem[\protect\citeauthoryear{{Kalirai}}{{Kalirai}}{2012}]{2012Natur.488..684K}
{Kalirai} J.~S.,  2012, \mn@doi [\nat] {10.1038/nature11343}, \href
  {http://adsabs.harvard.edu/abs/2012Natur.488..684K} {488, 684}

\bibitem[\protect\citeauthoryear{{Kalirai}, {Saul Davis}, {Richer}, {Bergeron},
  {Catelan}, {Hansen}  \& {Rich}}{{Kalirai} et~al.}{2009}]{2009ApJ...705..408K}
{Kalirai} J.~S.,  {Saul Davis} D.,  {Richer} H.~B.,  {Bergeron} P.,  {Catelan}
  M.,  {Hansen} B.~M.~S.,   {Rich} R.~M.,  2009, \mn@doi [\apj]
  {10.1088/0004-637X/705/1/408}, \href
  {http://adsabs.harvard.edu/abs/2009ApJ...705..408K} {705, 408}

\bibitem[\protect\citeauthoryear{Kalirai et~al.,}{Kalirai
  et~al.}{2012}]{1538-3881-143-1-11}
Kalirai J.~S.,  et~al., 2012, \aj, 143, 11

\bibitem[\protect\citeauthoryear{{Kilic}, {Munn}, {Harris}, {von Hippel},
  {Liebert}, {Williams}, {Jeffery}  \& {DeGennaro}}{{Kilic}
  et~al.}{2017}]{2017ApJ...837..162K}
{Kilic} M.,  {Munn} J.~A.,  {Harris} H.~C.,  {von Hippel} T.,  {Liebert} J.~W.,
   {Williams} K.~A.,  {Jeffery} E.,   {DeGennaro} S.,  2017, \mn@doi [\apj]
  {10.3847/1538-4357/aa62a5}, \href
  {http://adsabs.harvard.edu/abs/2017ApJ...837..162K} {837, 162}

\bibitem[\protect\citeauthoryear{{Kowalski} \& {Saumon}}{{Kowalski} \&
  {Saumon}}{2006}]{2006ApJ...651L.137K}
{Kowalski} P.~M.,  {Saumon} D.,  2006, \mn@doi [\apjl] {10.1086/509723}, \href
  {http://adsabs.harvard.edu/abs/2006ApJ...651L.137K} {651, L137}

\bibitem[\protect\citeauthoryear{{Metcalfe}, {Montgomery}  \&
  {Kanaan}}{{Metcalfe} et~al.}{2004}]{2004ApJ...605L.133M}
{Metcalfe} T.~S.,  {Montgomery} M.~H.,   {Kanaan} A.,  2004, \mn@doi [\apjl]
  {10.1086/420884}, \href {http://adsabs.harvard.edu/abs/2004ApJ...605L.133M}
  {605, L133}

\bibitem[\protect\citeauthoryear{{Moehler}, {Koester}, {Zoccali}, {Ferraro},
  {Heber}, {Napiwotzki}  \& {Renzini}}{{Moehler}
  et~al.}{2004}]{2004A&A...420..515M}
{Moehler} S.,  {Koester} D.,  {Zoccali} M.,  {Ferraro} F.~R.,  {Heber} U.,
  {Napiwotzki} R.,   {Renzini} A.,  2004, \mn@doi [\aap]
  {10.1051/0004-6361:20035819}, \href
  {http://adsabs.harvard.edu/abs/2004A%26A...420..515M} {420, 515}

\bibitem[\protect\citeauthoryear{{Munn} et~al.,}{{Munn}
  et~al.}{2017}]{2017AJ....153...10M}
{Munn} J.~A.,  et~al., 2017, \mn@doi [\aj] {10.3847/1538-3881/153/1/10}, \href
  {http://adsabs.harvard.edu/abs/2017AJ....153...10M} {153, 10}

\bibitem[\protect\citeauthoryear{{Paxton}, {Bildsten}, {Dotter}, {Herwig},
  {Lesaffre}  \& {Timmes}}{{Paxton} et~al.}{2011}]{2011ApJS..192....3P}
{Paxton} B.,  {Bildsten} L.,  {Dotter} A.,  {Herwig} F.,  {Lesaffre} P.,
  {Timmes} F.,  2011, \mn@doi [\apjs] {10.1088/0067-0049/192/1/3}, \href
  {http://adsabs.harvard.edu/abs/2011ApJS..192....3P} {192, 3}

\bibitem[\protect\citeauthoryear{{Potekhin} \& {Chabrier}}{{Potekhin} \&
  {Chabrier}}{2010}]{2010CoPP...50...82P}
{Potekhin} A.~Y.,  {Chabrier} G.,  2010, \mn@doi [Contributions to Plasma
  Physics] {10.1002/ctpp.201010017}, \href
  {http://adsabs.harvard.edu/abs/2010CoPP...50...82P} {50, 82}

\bibitem[\protect\citeauthoryear{{Reimers}}{{Reimers}}{1975}]{1975MSRSL...8..369R}
{Reimers} D.,  1975, Memoires of the Societe Royale des Sciences de Liege,
  \href {http://adsabs.harvard.edu/abs/1975MSRSL...8..369R} {8, 369}

\bibitem[\protect\citeauthoryear{Richer, Heyl, Anderson, Kalirai, Shara,
  Fahlman  \& Rich}{Richer et~al.}{2013a}]{Rich1347tuc}
Richer H.,  Heyl J.,  Anderson J.,  Kalirai J.~S.,  Shara M.,  Fahlman G.,
  Rich R.~M.,  2013a, \apjl, 771, L15

\bibitem[\protect\citeauthoryear{Richer et~al.,}{Richer
  et~al.}{2013b}]{2013ApJ...778..104R}
Richer H.~B.,  et~al., 2013b, \mn@doi [\apj] {10.1088/0004-637X/778/2/104},
  \href {http://adsabs.harvard.edu/abs/2013ApJ...778..104R} {778, 104}

\bibitem[\protect\citeauthoryear{{Salaris}, {Held}, {Ortolani}, {Gullieuszik}
  \& {Momany}}{{Salaris} et~al.}{2007}]{2007A&A...476..243S}
{Salaris} M.,  {Held} E.~V.,  {Ortolani} S.,  {Gullieuszik} M.,   {Momany} Y.,
  2007, \mn@doi [\aap] {10.1051/0004-6361:20078445}, \href
  {http://adsabs.harvard.edu/abs/2007A%26A...476..243S} {476, 243}

\bibitem[\protect\citeauthoryear{{Sirianni} et~al.}{{Sirianni}
  et~al.}{2005}]{2005PASP..117.1049S}
{Sirianni} M.,  et~al., 2005, \mn@doi [\pasp] {10.1086/444553}, \href
  {http://adsabs.harvard.edu/abs/2005PASP..117.1049S} {117, 1049}

\bibitem[\protect\citeauthoryear{{Tassoul}, {Fontaine}  \& {Winget}}{{Tassoul}
  et~al.}{1990}]{1990ApJS...72..335T}
{Tassoul} M.,  {Fontaine} G.,   {Winget} D.~E.,  1990, \mn@doi [\apjs]
  {10.1086/191420}, \href {http://adsabs.harvard.edu/abs/1990ApJS...72..335T}
  {72, 335}

\bibitem[\protect\citeauthoryear{{Tremblay}, {Bergeron}  \&
  {Gianninas}}{{Tremblay} et~al.}{2011}]{2011ApJ...730..128T}
{Tremblay} P.-E.,  {Bergeron} P.,   {Gianninas} A.,  2011, \mn@doi [\apj]
  {10.1088/0004-637X/730/2/128}, \href
  {http://adsabs.harvard.edu/abs/2011ApJ...730..128T} {730, 128}

\bibitem[\protect\citeauthoryear{{Tremblay}, {Ludwig}, {Freytag}, {Fontaine},
  {Steffen}  \& {Brassard}}{{Tremblay} et~al.}{2015}]{2015ApJ...799..142T}
{Tremblay} P.-E.,  {Ludwig} H.-G.,  {Freytag} B.,  {Fontaine} G.,  {Steffen}
  M.,   {Brassard} P.,  2015, \mn@doi [\apj] {10.1088/0004-637X/799/2/142},
  \href {http://adsabs.harvard.edu/abs/2015ApJ...799..142T} {799, 142}

\bibitem[\protect\citeauthoryear{{Winget}}{{Winget}}{1995}]{1995BaltA...4..129W}
{Winget} D.~E.,  1995, Baltic Astronomy, \href
  {http://adsabs.harvard.edu/abs/1995BaltA...4..129W} {4, 129}

\bibitem[\protect\citeauthoryear{{Winget}, {Kepler}, {Campos}, {Montgomery},
  {Girardi}, {Bergeron}  \& {Williams}}{{Winget}
  et~al.}{2009}]{2009ApJ...693L...6W}
{Winget} D.~E.,  {Kepler} S.~O.,  {Campos} F.,  {Montgomery} M.~H.,  {Girardi}
  L.,  {Bergeron} P.,   {Williams} K.,  2009, \mn@doi [\apjl]
  {10.1088/0004-637X/693/1/L6}, \href
  {http://adsabs.harvard.edu/abs/2009ApJ...693L...6W} {693, L6}

\bibitem[\protect\citeauthoryear{{Woodley} et~al.,}{{Woodley}
  et~al.}{2012}]{2012AJ....143...50W}
{Woodley} K.~A.,  et~al., 2012, \mn@doi [\aj] {10.1088/0004-6256/143/2/50},
  \href {http://adsabs.harvard.edu/abs/2012AJ....143...50W} {143, 50}

\bibitem[\protect\citeauthoryear{{van Horn}}{{van
  Horn}}{1967}]{1967AJ.....72R.834V}
{van Horn} H.~M.,  1967, \mn@doi [\aj] {10.1086/110522}, \href
  {http://adsabs.harvard.edu/abs/1967AJ.....72R.834V} {72, 834}

\bibitem[\protect\citeauthoryear{{van Horn}}{{van
  Horn}}{1968}]{1968ApJ...151..227V}
{van Horn} H.~M.,  1968, \mn@doi [\apj] {10.1086/149432}, \href
  {http://adsabs.harvard.edu/abs/1968ApJ...151..227V} {151, 227}

\makeatother
\end{thebibliography}
\bibliographystyle{mnras}


\bsp	
\label{lastpage}
\end{document}